\documentclass[ aip, jmp, amsmath,amssymb, reprint, ]{revtex4-1}

\usepackage{amssymb}
\usepackage{amsfonts}
\usepackage{amsmath}
\usepackage{amsthm}
\usepackage{epsfig}
\usepackage{color} 
\usepackage{subfigure}

\graphicspath{{./}{./Figures/}}

\usepackage{latexsym}
\usepackage{mathrsfs}
\usepackage{bm}
\usepackage{comment}

\newcommand{\m}[1]{\begin{pmatrix}#1\end{pmatrix}}

\newcommand{\matr}[1]{{{\bm{#1}}}}    
\renewcommand{\vec}[1]{{\bm{#1}}}    

\begin{document}

\title{Chimera states in hierarchical networks of Van der Pol oscillators} 

\author{Stefan Ulonska}
\email[]{stefan.ulonska@fu-berlin.de}
\affiliation{Institut f\"ur Theoretische Physik, Technische Universit\"at Berlin, Hardenbergstra{\ss}e 36, 10623 Berlin, Germany}

\author{Iryna Omelchenko}
\affiliation{Institut f\"ur Theoretische Physik, Technische Universit\"at Berlin, Hardenbergstra{\ss}e 36, 10623 Berlin, Germany}

\author{Anna Zakharova}
\affiliation{Institut f\"ur Theoretische Physik, Technische Universit\"at Berlin, Hardenbergstra{\ss}e 36, 10623 Berlin, Germany}

\author{Eckehard Sch\"oll}
\email[]{schoell@physik.tu-berlin.de}
\affiliation{Institut f\"ur Theoretische Physik, Technische Universit\"at Berlin, Hardenbergstra{\ss}e 36, 10623 Berlin, Germany}

\date{\today}

\begin{abstract}

Chimera states are complex spatio-temporal patterns that consist of coexisting domains of coherent and incoherent dynamics. We analyse chimera states in networks of Van der Pol oscillators with hierarchical coupling topology. We investigate the stepwise transition from a nonlocal to a hierarchical topology, and propose the network clustering coefficient as a measure to establish a link between the existence of chimera states and the compactness of the initial base pattern of a hierarchical topology; we show that a large clustering coefficient promotes the occurrence of chimeras.
Depending on the level of hierarchy and base pattern, we obtain chimera states with different numbers of incoherent domains. We investigate the chimera regimes as a function of coupling strength and nonlinearity parameter of the individual oscillators. 
The analysis of a network with larger base pattern resulting in larger clustering coefficient reveals two different types of chimera states and highlights the increasing role of amplitude dynamics. 
\end{abstract}


\maketitle

\begin{quotation}
Chimera states are an example of intriguing partial synchronization patterns appearing in networks of identical oscillators with complex coupling schemes. They exhibit a hybrid structure combining coexisting spatial domains of coherent (synchronized) and incoherent (desynchronized) dynamics, and were first reported for the model of phase oscillators\cite{KUR02a,ABR04}.
Recent studies have demonstrated the emergence of chimera states in a variety of topologies, and for different types of individual dynamics.
In this paper, the transition from nonlocal to hierarchical (quasi-fractal) topologies is studied systematically in a network of identical Van der Pol oscillators. The clustering coefficient and symmetry properties are used to classify different topologies with respect to the occurrence of chimera states. We show that symmetric topologies with large clustering coefficients promote the emergence of chimera states, while they are suppressed by slight topological asymmetries or small clustering coefficient.
\end{quotation}
\section{Introduction}
\label{sec:intro}

The analysis of coupled oscillatory systems is an important research field bridging between nonlinear dynamics, network science, and
statistical physics, with a variety of applications in physics, biology, and technology~\cite{PIK01,BOC06a}. The study of large networks with complex coupling schemes continues to open up new unexpected dynamical scenarios. 

The last decade has seen an increasing interest in chimera states in dynamical
networks \cite{LAI09,MOT10,OME10a,OME12a,MAR10b,WOL11a,BOU14,PAN15,SCH16b}. First obtained in the systems of
phase oscillators, chimeras can also be found in a large variety of different systems including time-discrete maps~\cite{OME11,SEM15a},
time-continuous chaotic models~\cite{OME12}, neural systems~\cite{OME13,HIZ13,VUE14a,TSI16}, Boolean networks
\cite{ROS14a}, population dynamics \cite{HIZ15}, Van der Pol oscillators \cite{OME15a}, and quantum oscillator systems \cite{BAS15}.
Moreover, chimera states allow for higher spatial dimensions~\cite{OME12a,SHI04,MAR10,PAN13,PAN15,PAN15a,MAI15}.
Together with the initially reported chimera states, which consist of one coherent and one incoherent domain, new types
of these peculiar states having multiple incoherent regions~\cite{SET08,OME13,MAI14,XIE14,VUE14a,OME15a}, as well as
amplitude-mediated~\cite{SET13,SET14}, and pure amplitude chimera and chimera death states~\cite{ZAK14} were discovered.

In many systems, the form of the coupling defines the possibility to obtain chimera states. The nonlocal coupling has
generally been assumed to be a necessary condition for chimera states to evolve in coupled systems. However, recent
studies have shown that even global all-to-all coupling~\cite{SET14,YEL14,SCH14g,BOE15}, as well as more complex
coupling topologies allow for the existence of chimera states~\cite{KO08,SHA10,LAI12,YAO13,ZHU14,OME15,HIZ15,TSI16}. Furthermore,
time-varying network structures can give rise to alternating chimera states \cite{BUS15}.
Chimera states have also been shown to be robust against inhomogeneities of the local dynamics and coupling topology~\cite{LAI10,OME15},
as well as against noise \cite{LOO16}, or they might even be induced by noise \cite{SEM16,SEM15b}.

Possible applications of chimera states in natural and technological systems include the phenomenon of unihemispheric sleep~\cite{RAT00}, bump states in neural systems~\cite{LAI01,SAK06a}, epileptic seizures~\cite{ROT14}, power grids~\cite{FIL08}, or social systems~\cite{GON14}.  
Many works considering chimera states have mostly been based on numerical results. A deeper bifurcation analysis~\cite{OME13a} and even a possibility to control chimera states~\cite{SIE14c,BIC15,OME16} were obtained only recently.

The experimental verification of chimera states was first demonstrated in optical~\cite{HAG12} and chemical~\cite{TIN12,NKO13} systems. Further experiments involved  mechanical~\cite{MAR13,KAP14}, electronic~\cite{LAR13,GAM14}, optoelectronic delayed-feedback \cite{LAR15} and electrochemical~\cite{WIC13,WIC14,SCH14a} oscillator systems, Boolean networks~\cite{ROS14a}, and optical combs \cite{VIK14}.

\begin{figure*}[t]%
\includegraphics[width=0.8\textwidth]{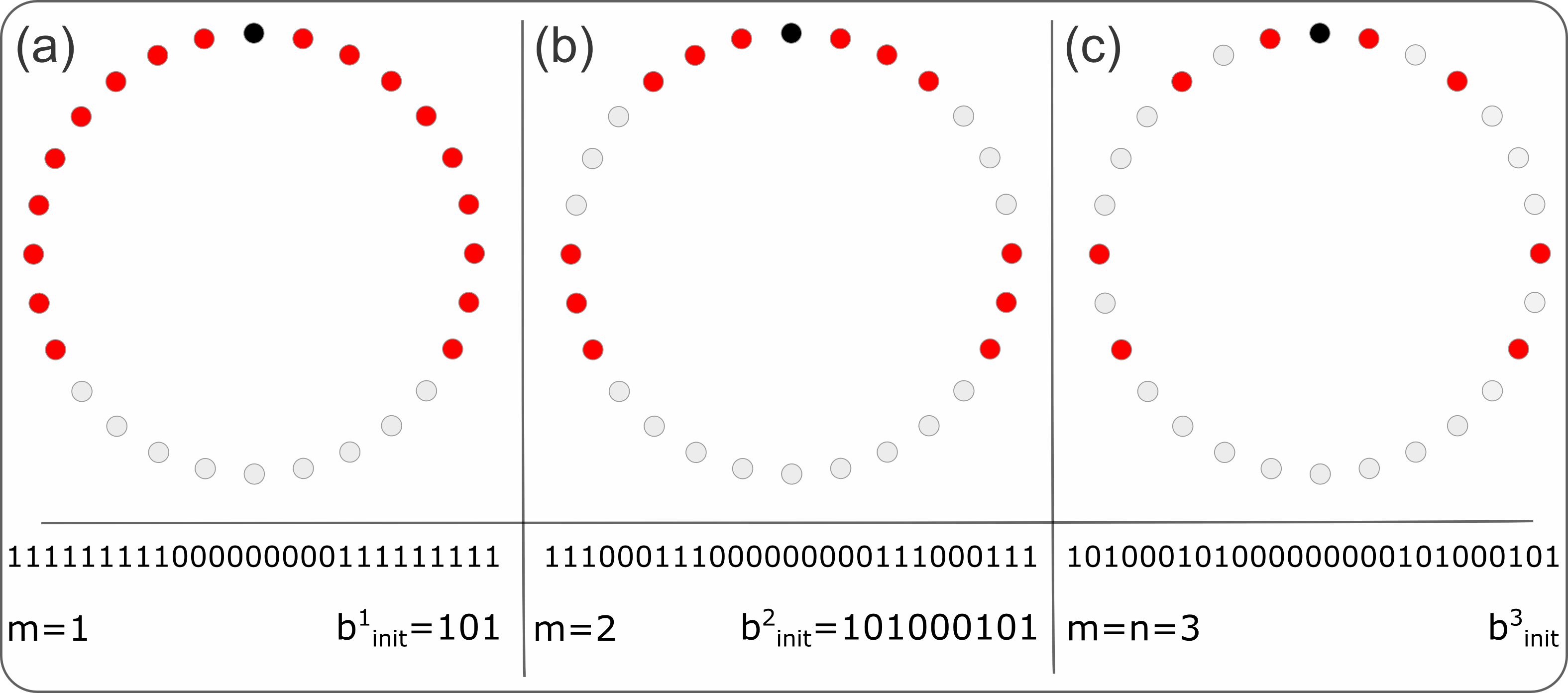}%
\caption{Transition from nonlocal to hierarchical coupling via hierarchical steps m. The reference node is coloured in black, linked nodes in red and unconnected nodes (gaps) in gray. The initial base pattern of all panels is $b_{init}=(101)$, the level of hierarchy is $n=3$, $N=b^n+1=28$ nodes. (a)~$m=1$, each element in the initial base pattern is expanded by $\frac{27}{3}=9$ elements and the final 1-step hierarchical system corresponds to nonlocal coupling where each element is coupled to its $R=9$ nearest neighbors in both directions. Clustering coefficient $C(101,3,1)=0.705882$ and link density $\rho=0.64$; (b)~$m=2$, expansion by $\frac{27}{9}=3$ elements to a 2-step hierarchical system. $C(101,3,2)=0.409091$ and $\rho=0.428$; (c)~$m=n=3$, fully hierarchical or $n-$step hierarchical system without further expansion of the base pattern. $C(101,3,3)=0$ and $\rho=0.286$. With each $m-$step, the total number of links for each node and the clustering coefficient decrease.}%
\label{fig:vdpmod2}%
\end{figure*}
Recent results in the area of neuroscience increased the interest in irregular coupling topologies. 
Diffusion Tensor Magnetic Resonance Imaging (DT-MRI) studies revealed an intricate
architecture in the neuron interconnectivity of the human and mammalian brain: the connectivity of the neuron axons network represents a hierarchical (quasi-fractal)
geometry with fractal dimensions varying between $2.3$ and $2.8$, depending 
on the local properties, on the subject, and on the noise reduction
threshold \cite{KAT09,EXP11,KAT12,KAT12a,PRO12}.

Motivated by these studies, the goal of the present manuscript is to analyze different networks with hierarchical connectivity and systematically explore the mechanisms of formation of chimera states in such networks. We aim to uncover characteristic measures of the
hierarchical connectivities which determine the emergence of chimeras. As a key measure we identify the clustering coefficient.
We focus on the stepwise transition of network topology from nonlocal to hierarchical, and analyze different types of chimera states that can arise in systems with different numbers of hierarchical steps. 
The dynamics of individual units in the network is governed by the Van der Pol oscillator~\cite{POL26}, which has a long history of
being used in both the physical and biological sciences, and allows for a continuous transition between sinusoidal and strongly nonlinear relaxation oscillations by tuning a single parameter.

\section{The model}
\label{sec:model}
In our study, we consider a system of $N$ identical Van der Pol oscillators with varying ring topologies, which are given by the respective adjacency matrices $\matr{G}$. The dynamical equations for the 2-dimensional phase space variable $\vec{x}_k=(u_k, \dot{u}_k)^T=(u_k, v_k)^T \in \mathbb{R}^2$ are:
\begin{align}
\vec{\dot{x}}_i(t) &=  \vec{F}(\vec{x}_i(t)) +  \frac{\sigma}{g} \sum^{N}_{j=1}G_{ij} \matr{H}(\vec{x}_j-\vec{x}_i)
\label{eqn:gen1}
\end{align}
with $i \in \{1,...,N\}$. The dynamics of each individual oscillator is governed by 
\begin{eqnarray}
\label{eq:localdyn}
\vec{F}(\vec{x})=
\left(\!
\begin{array}{*{1}{c}}
v\\
\varepsilon(1-u^2)v-u
\end{array}
\!\right) ,
\end{eqnarray}
where  $\varepsilon$ denotes the bifurcation parameter.
The uncoupled Van der Pol oscillator has a stable fixed point at $\vec{x}=0$ for $\varepsilon<0$ and undergoes an Andronov-Hopf bifurcation at $\varepsilon=0$. Here, only $\varepsilon>0$ is considered.  The parameter~$\sigma$ denotes the coupling strength,  and $g=\sum^{N}_{j=1}G_{ij}$ is the number of links for each node (corresponding to the row sum of $\matr{G}$).  The local interaction is realized as diffusive coupling with local coupling matrix $\matr{H}=\m{0&0\\b_1&b_2}$ and real interaction parameters $b_1$ and $b_2$. In accordance with Omelchenko et al.~\cite{OME15a}, throughout the manuscript we fix parameters $b_1=1.0$ and $b_2=0.1$, which allows us to observe chimera states in nonlocally coupled systems.

\subsection{An algorithm to construct hierarchical topologies}
Hierarchical topologies can be generated using a classical Cantor construction algorithm for a fractal set\cite{MAN83,FED88}. This iterative hierarchical procedure starts from a \emph{base pattern} or initiation string $b_{init}$ of length $b$, where each element represents either a link ('$1$') or a gap ('$0$'). The number of links contained in $b_{init}$ is referred to as $c_1$. In each iterative step, each link is replaced by the initial base pattern, while each gap is replaced by $b$ gaps. Thus, each iteration increases the size of the final bit pattern, such that after $n$ iterations the total length is $N=b^n$. Since the hierarchy is truncated at a finite $n$, we call the resulting pattern hierarchical rather than fractal. 
Using the resulting string as the fundamental row of the circulant adjacency matrix $\matr{G}$, a ring network of $N=b^n$ nodes with hierarchical topology is generated \cite{OME15,HIZ15,TSI16}.  Here we will slightly modify this procedure by adding an additional zero in the first instance of the sequence, which corresponds to the self-coupling. Note that the diffusive coupling scheme in Eq.(\ref{eqn:gen1}) cancels out any instantaneous self-coupling. Therefore there is no net effect of the diagonal elements of the adjacency matrix $G_{ii}$ on the network dynamics, and hence the first link in the clockwise sense from the reference node is effectively removed from the link pattern. Without our modification, this would lead to a breaking of the base pattern symmetry, i.e., if the base pattern is symmetric, the resulting coupling topology would not be so, since the first link to the right is missing from the final link pattern.
Our procedure, in contrast, ensures the preservation of an initial symmetry of $b_{init}$ in the final link pattern, which is crucial for the observation of chimera states, since asymmetric coupling leads to a drift of the chimera \cite{OME16}. Thus a ring network of $N=b^n+1$ nodes is generated.

While fully hierarchical topologies can be generated using this classical construction algorithm, a further modification allows us to study systematically the transition between nonlocal and hierarchical topologies via a stepwise iteration process. Nonlocal coupling schemes have widely been used in the context of chimera state research~\cite{OME15a} and do therefore provide a good reference point to compare hierarchical networks to.

A nonlocal topology can be generated from a base pattern $b_{init}$ of length $b$, which contains an equal number of links only at its beginning and end (for instance ($101$) or ($110011$)). The link pattern is then expanded to a predetermined system size $N=b^n+1$, corresponding to the final size of a fully hierarchical topology, by replacing each element with $\frac{N-1}{b}$ copies of itself and again adding the additional zero in first instance of the final pattern. Thus, a suitable base pattern $b_{init}$ can be used to either construct a fully hierarchical or a nonlocal topology. The stepwise transition between these two types of topologies is realized as follows: First, $b_{init}$ is iterated $m$ times according to the Cantor construction process, generating a pattern of size $b^m$. Afterwards, this pattern is expanded to the final size $N$ by replacing each element with $\frac{N-1}{b^m}$ copies of itself. The initial base pattern $b_{init}=(101)$ and a predetermined system size of $N=27+1$ $(b=3, n=3)$ provide a 
simple example. The resulting coupling topologies are illustrated in Fig.~\ref{fig:vdpmod2}. The number of Cantor iterations of $b$ \emph{before} the expansion is defined as the the $m^{th}$ hierarchical step, with $m \in \{1,2,...,n\}$ such that
\begin{itemize}
	\item[$m=1$:] $b_{init}^1$=($101$), the initial base pattern is expanded to a 1-step hierarchical topology by replacing each element with $\frac{27}{3}=9$ copies of itself. This corresponds to nonlocal coupling with coupling radius $r=\frac{R}{N}=\frac{9}{27}=0.333$. See Fig.~\ref{fig:vdpmod2}(a).
	\item[$m=2$:] $b_{init}^2$=($101000101$), the once iterated initial base pattern is expanded to a $2-$step hierarchical topology by replacing each element with $\frac{27}{9}=3$ copies of itself. This marks the first step in the transition from a nonlocal $(m=1)$ to a fully hierarchical $(m=n=3)$ topology. See Fig.~\ref{fig:vdpmod2}(b).
	\item[$m=n=3$:] $b_{init}^3$=($101000101000000000101000101$), this link pattern is of size $N$ for $m=n=3$ and corresponds to the fully hierarchical or $n-$step hierarchical topology. See Fig.~\ref{fig:vdpmod2}(c).
\end{itemize}
Thus, taking up to $m=n$ steps in the hierarchical expansion it is possible to tune a suitable initial base pattern between a nonlocal and a fully hierarchical topology. In the following, an $m-$step hierarchical topology is denoted as $(b_{init})^n$, where $b_{init}$ is the underlying base pattern, $n$ the level of hierarchy and $m$ the hierarchical step in a transition topology. 
\subsection{Clustering coefficients }
\label{ssec:clusterc}
\begin{figure}[t]%
\includegraphics[width=1.0\columnwidth]{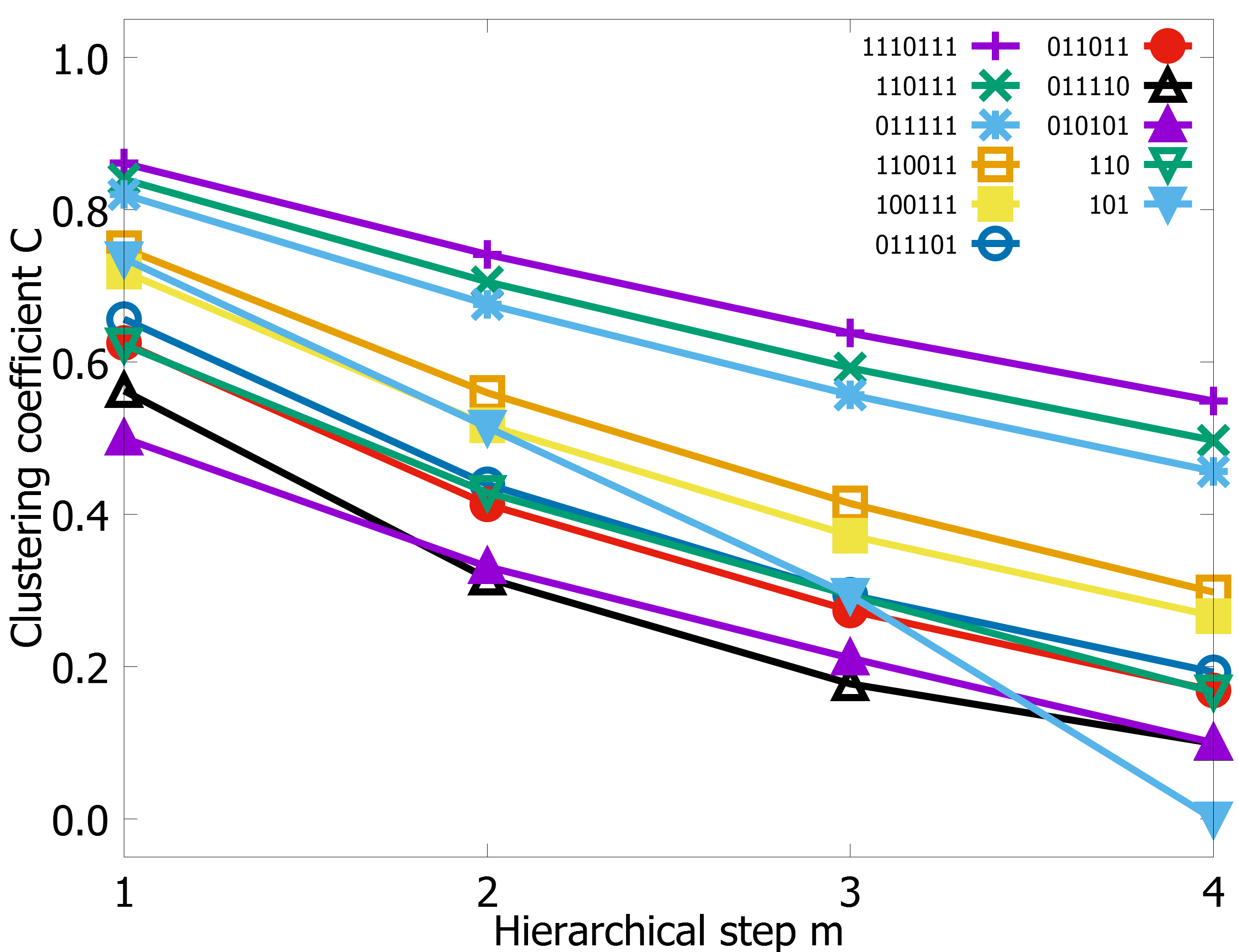}%
\caption{Clustering coefficients $C$ for networks with topologies varied stepwise from nonlocal to hierarchical for different base patterns (see legend). The hierarchical step $m\in \{1,...,n=4\}$ is used to tune between nonlocal coupling and fully hierarchical coupling. At higher hierarchical step $m$ the clustering coefficients decreases due to the disproportional increase of gaps versus links in the base pattern. Non-compact base patterns lead to vanishing clustering coefficients $C$. This suppresses chimera states in hierarchical networks generated from these base patterns. System sizes (n=4): $N=82$ for $b=3$, $N=1297$ for $b=6$, $N=2402$ for $b=7$.}%
\label{fig:cc1}%
\end{figure}
Besides the bifurcation parameter $\varepsilon$ and the coupling strength $\sigma$, the topological quantities $b_{init}$, $c_1$, $n$ and the resulting link density $\rho=\frac{c_1^n}{N}$ and fractal dimension $d_f=\ln c_1/\ln b$ are important parameters in the study of chimera states in hierarchical systems. However, since there are several distributions of links for a given set of $b$ and $c_1$ that result in unique topological structures, the arrangement of links in $b_{init}$ has to be accounted for. Therefore, we propose to consider the local clustering coefficient introduced by Watts and Strogatz~\cite{WAT98}, which, for a directed graph \textbf{G}=$(V,E)$ containing a set of nodes $V$ and edges $E$, describes the number of links in the neighbourhood $N_i = \{v_j : e_{ij} \in E \lor e_{ji} \in E\}$ relative to the maximum number of links possible. Since a directed graph distinguishes between $e_{ij} $ and $e_{ji}$, the maximum number of links is given by $k_i\cdot(k_i-1)$ and the clustering coefficient $C_i$ for a node $v_i$ is defined as
\begin{align}
C_i &= \frac{|\{e_{jk}: v_j,v_k \in N_i, e_{jk} \in E\}|}{k_i(k_i-1)}
\label{eqn:cc1}
\end{align}
\begin{figure*}[t]%
\includegraphics[width=0.9\textwidth]{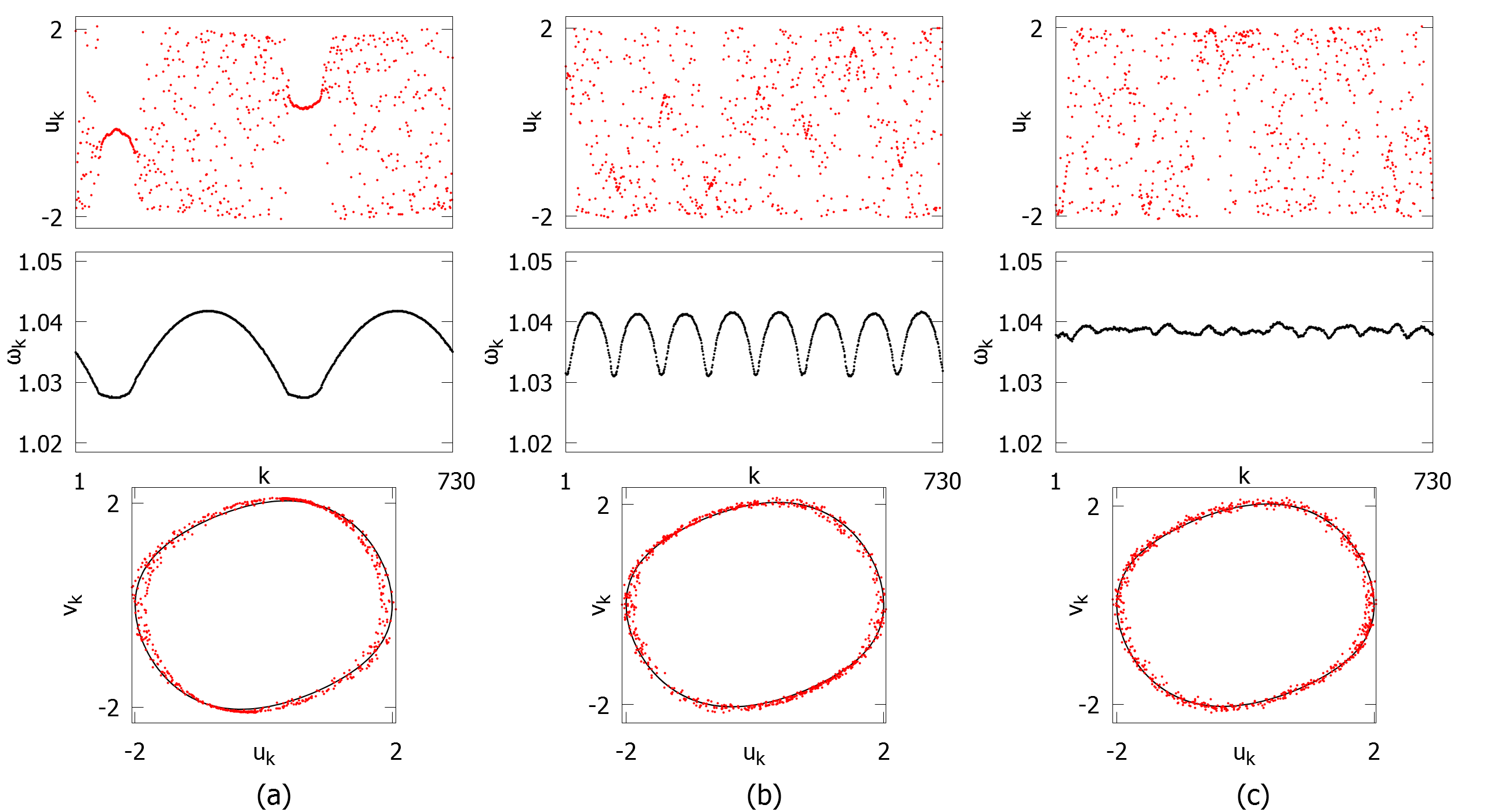}
\caption{Chimera states in transiting topology with $b_{init}=(101)$, $n=6$, $N=730$, $\sigma=0.09$, $\varepsilon=0.2$. Snapshots of variables $u_k$ (upper panels), mean phase velocities $\omega_k$ (middle panels), and snapshots in the phase space $(u_k,v_k)$ (bottom panels, limit cycle of the uncoupled unit shown in black). (a)~Hierarchical step $m=1$, corresponding to nonlocal coupling with $r=0.333$, 2-chimera state, clustering coefficient $C(101,6,1)=0.748451$ and link density $\rho=0.667$;  (b)~$m=2$, 8-chimera state, $C(101,6,2)=0.55727$ and $\rho=0.444$; (c)~$m=3$, completely incoherent state, $C(101,6,3)=0.408141$ and $\rho=0.296$. No chimeras are observed for further steps in hierarchical topology. Initial conditions as in \cite{OME15a}, Fig.1(b).}%
\label{fig:chimdec1}%
\end{figure*}
While common nonlocal coupling schemes can rely on the coupling range (or variations thereof \cite{OME15}) as a definite measure, this is not the case in hierarchically coupled networks. Different arrangement of links change the compactness of the base pattern and the final compactness of the system strongly depends on the distribution properties of the $c_1$ links in a base pattern $b_{init}$. The clustering coefficient is such a suitable measure since it directly expresses the compactness of links in the final system. In the following $C(b_{init},n,m)$ denotes the clustering coefficient of an $m-$step hierarchical topology with base pattern $b_{init}$.
In the transition scenario from nonlocal to hierarchical topologies, the system size and the symmetry properties stay the same throughout. However, each hierarchical step modifies the compactness and total number of links of the final topology which changes the clustering coefficient $C$ as well as the link density $\rho$. Since hierarchical coupling introduces irregular and long-ranging links and gaps, the clustering coefficient will decrease when transiting towards a hierarchical topology.\\
 Out of all the $2^b$ possible base patterns for a given length $b$, only a fraction is relevant for discussion in the context of hierarchical networks. For example, mirror symmetric base patterns (such as ($110101$) and ($101011$)) result in topologies with identical clustering coefficients, while patterns containing only one link would result in a network with only a single link to each node.  In the case of $b=6$, out of $64$ possible patterns, only $31$ will yield unique hierarchical topologies. Fig.~\ref{fig:cc1} demonstrates clustering coefficients $C(b_{init},n,m)$ as a function of the hierarchical step $m$ for a selected number of representative base patterns.

\begin{figure*}[t]%
\includegraphics[width=0.9\textwidth]{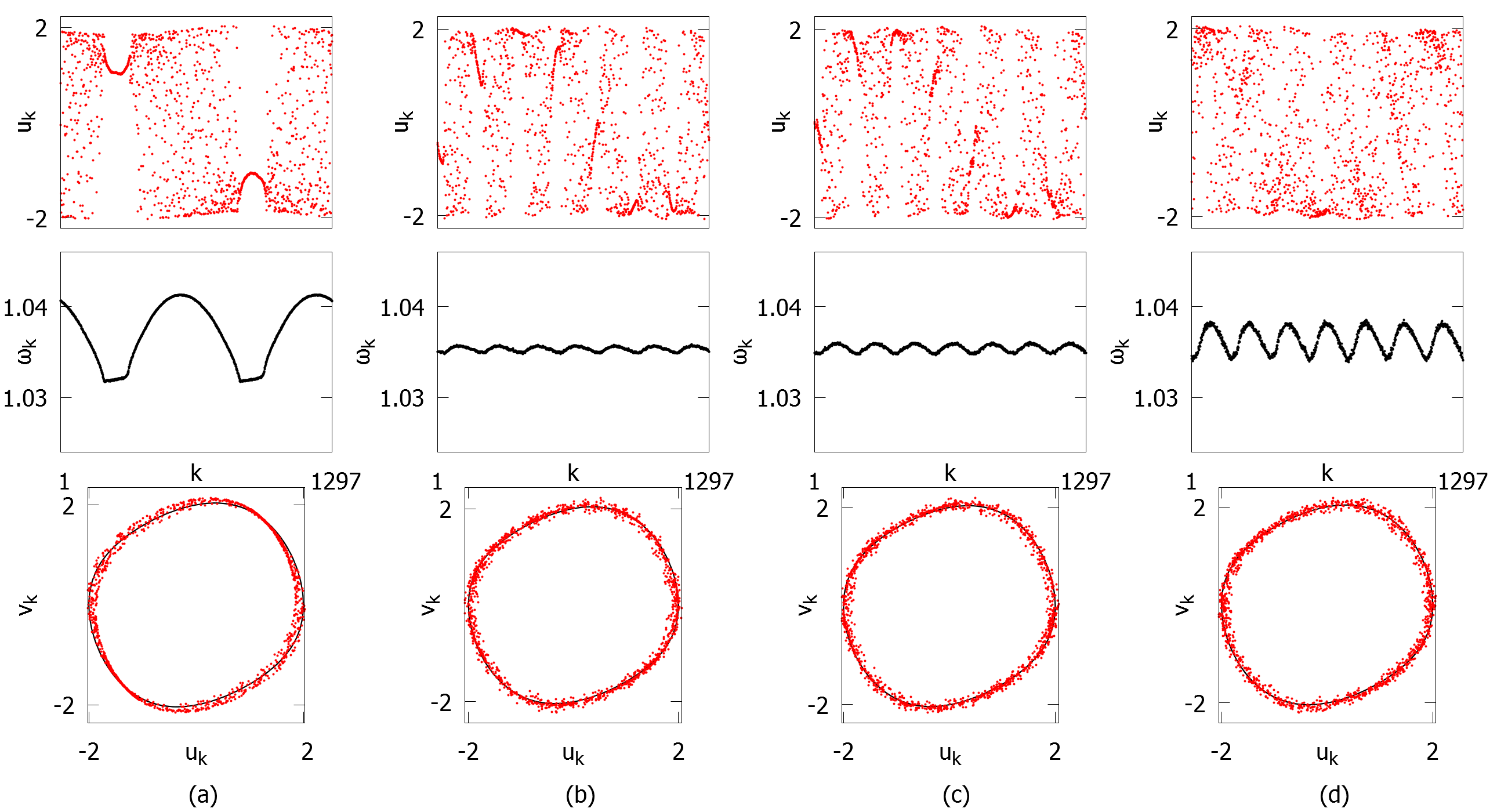}
\caption{Chimera states in transiting topology with $b_{init}=(110011)$, $n=4$, $N=1297$, $\sigma=0.09$, $\varepsilon=0.2$, random initial conditions. Snapshots of variables $u_k$ (upper panels), mean phase velocities $\omega_k$ (middle panels), and snapshots in the phase space $(u_k,v_k)$ (bottom panels, limit cycle of the uncoupled element shown in black). (a)~$m=1$, corresponding to nonlocal coupling with $r=0.256$, $C(110011,4,1)=0.749142$, $\rho=0.667$, $2-$chimera state; (b)~$m=2$, $C(110011,4,2)=0.559569$, $\rho=0.444$, $7-$chimera; (c)~$m=3$, $C(110011,4,3)=0.414161$, $\rho=0.298$, $7-$ chimera remains; (d)~$m=n=4$, fully hierarchical network with $7-$chimera and more pronounced $\omega_k$ profile, $C(110011,4,4)=0.297791$ and $\rho=0.197$.}%
\label{fig:chimdec2}%
\end{figure*}
These results provide an instructive overview over the dependence of $C$ upon $m$. As expected, the clustering coefficients decrease across-the-board when introducing hierarchical components. With each step $m$, the number of gaps increases disproportionally compared to the number of links, since each gap in $b_{init}$ introduces $b$ new gaps with each iteration, while each link only leads to $c_1$ new links and adds $b-c_1$ further gaps. Thus, the link density decreases with each iteration of $m$ and the remaining links are distributed in a more irregular, hierarchical manner. A close analysis of $C$ with respect to the underlying base pattern shows that base patterns with gaps at either the end or the beginning of $b_{init}$ (such as ($011011$), Fig.~\ref{fig:cc1}, red dots) result in low clustering coefficients, compared to systems of equal link density but different distribution (such as ($110011$), orange squares). Furthermore, isolated links in $b_{init}$ (such as ($100111$), yellow squares) further decrease $C(b_{init},n=m)$ compared to 
topologies constructed from more compact base patterns (such as ($110011$), orange squares). However, more compact base patterns $b_{init}$ that have isolated links (such as ($011101$), blue circles) \emph{can have} larger $C(b_{init},n=m)$, as compared to topologies with same number of links but without isolated links (such as ($011011$), red dots). This can be explained by the fact that isolated links lead to far-ranging links rather than coupling to those nodes in their close neighbourhood. Considering the choice of suitable base patterns $b_{init}$ for the construction of hierarchical topologies which exhibit chimeras, it should be noted that base patterns with larger clustering coefficients are preferable. Therefore, base patterns should be chosen avoiding isolated links, gaps at the beginning or end, and with compact intervals, if possible. This highlights the basic trade-off when studying hierarchical topologies. Compactness is an important requirement for the existence of chimera states and at the same time, networks are less compact if they are more hierarchically structured. If one wants to study dynamic phenomena and maintain a hierarchical topology, this has to be balanced in a careful manner, since low $n$ decreases the actual degree of hierarchy of the system, while large $n$ produces highly hierarchical systems (fractal, in the case of $n \to \infty$) but with very low compactness. It will be shown in the following that hierarchical topologies resulting from base patterns that contain only isolated links for given $b$ and $c_1$ have the lowest clustering coefficient and hence the smallest probability of exhibiting chimera states.

\section{Chimera states in hierarchical topologies}
\label{sec:trans}
\begin{figure*}[t]%
\includegraphics[width=0.8\textwidth]{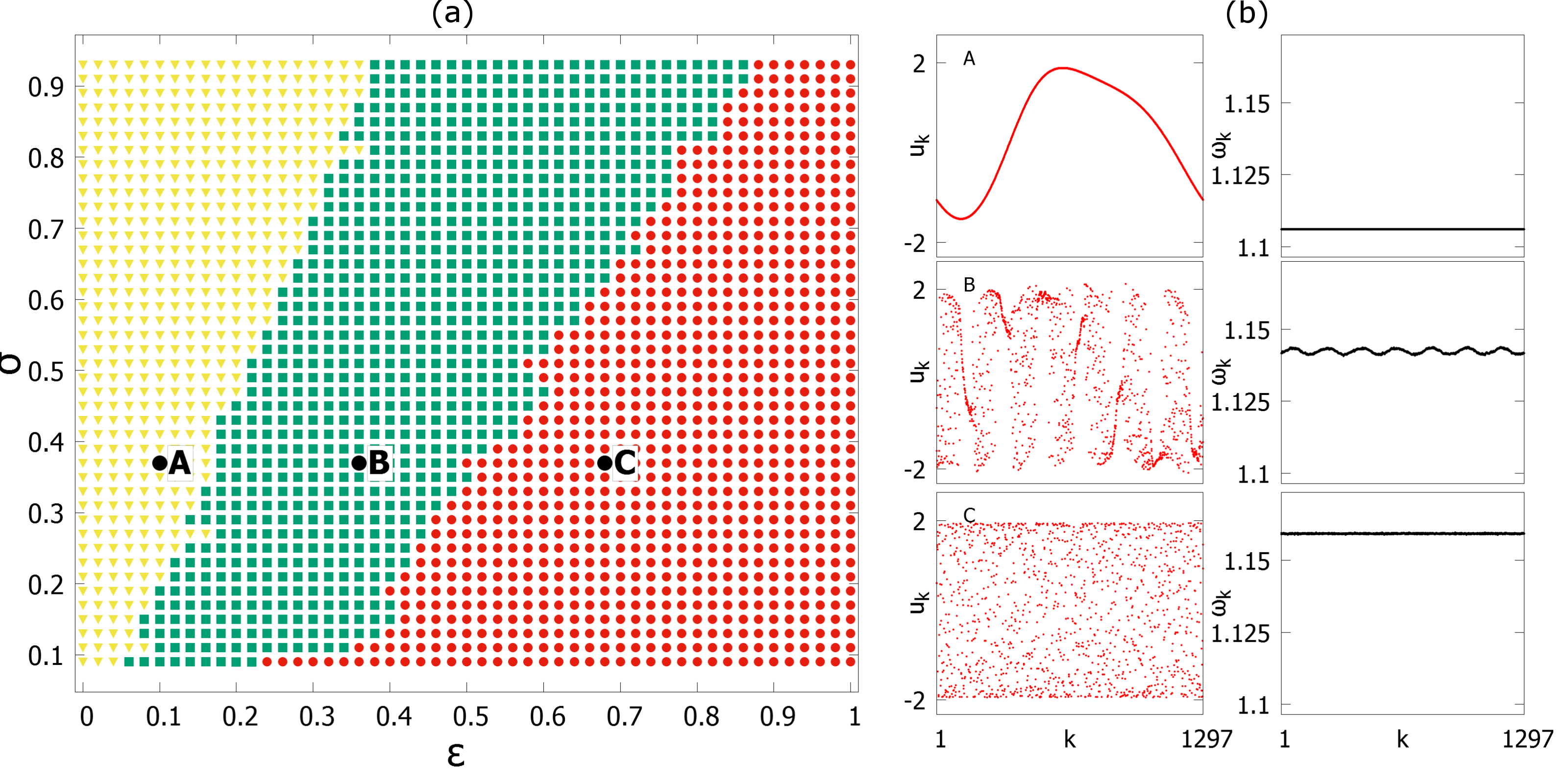}%
\caption{Stability regimes for the system $b_{init}=(110011)$, $n=4$ in an $m=3-$step hierarchical topology, $N=1297$, $C(110011,4,3)=0.414161$, $\rho=0.296$. (a) Diagram in the parameter space $(\varepsilon,\sigma)$: coherent state (yellow triangles),  $7-$chimera  state (green squares), incoherent state (red circles); (b) exemplary snapshots of variables~$u_k$ (left column) and mean phase velocities~$\omega_k$ (right column) for $\sigma=0.38$ and $\varepsilon=0.1$~(A), $\varepsilon=0.36$~(B), $\varepsilon=0.68$~(C).}%
\label{fig:dynm3}%
\end{figure*}
The simplest illustrative example of a hierarchical topology is constructed from the base pattern $(101)$. We consider this pattern in order to demonstrate the step-by-step transition from nonlocal to hierarchical network topology with $n=6$. At the initial step $m=1$, the system is a nonlocally coupled ring with $N=730$ and coupling radius $r=\frac{243}{730}=0.333$. The clustering coefficient for this topology is  $C(101,6,1)=0.748451$. This system has been studied in~\cite{OME15a}, where is was shown that depending on the coupling radius and strength, chimera states with different numbers of incoherent domains can be observed. Indeed, Fig.~\ref{fig:chimdec1}(a) shows a $2-$chimera state. Snapshots of the variable $u_k$ (upper panels) demonstrate a clear distinction between coherent and incoherent domains, snapshots in the phase space $(u_k,v_k)$ (bottom panels) show that the oscillators are scattered around the limit cycle of the uncoupled element which is shown in black.
The middle panels demonstrate the mean phase velocities for each oscillator calculated as $\omega_k=2\pi M_k/ \Delta T,$ $k=1,...,N$, where $M_k$ is the number of complete rotations around the origin performed by the $k-$th node during the time interval $\Delta T$. Throughout the paper, we use $\Delta T = 50 000$, which corresponds to several thousand periods.
Mean phase velocity profiles are widely used as a criterion to distinguish chimera states: constant $\omega_k$ corresponds to coherent domains, where neighbouring elements are phase-locked. Arc-like parts of the profiles correspond to incoherent domains.\\
Usually, chimera states with different numbers of incoherent domains exhibit high multistability~\cite{OME15a}, and the choice of initial condition is crucial. That is why we use specially prepared chimera-like initial conditions, as in~\cite{OME15a}, Fig.~1(b).\\
The $2-$step hierarchical network (Fig.~\ref{fig:chimdec1}(b)) is characterized by a smaller clustering coefficient $C(101,6,2)=0.557279$, and we observe an $8-$chimera state. This fact can be explained by drawing the analogy to nonlocally coupled systems, where a decrease of the coupling radius (the number of neighbours coupled to each element) results in an increase of the number of incoherent domains, i.e., the chimera multiplicity.

Already for the $3-$step hierarchical network shown in Fig.~\ref{fig:chimdec1}(c), the chimera state vanishes and is not observed for any further iterations up to $m=n=6$. This goes along with a decrease of the clustering coefficient to $C(101,6,3)=0.408141$ and further down to $C(101,6,6)=0$ for the fully hierarchical network (see Fig.~\ref{fig:cc1}). Transiting the system from a nonlocal to a hierarchical topology decreases the clustering coefficient as well as the total number of links to such an extent that chimera states do not occur after only $3$ out of $6$ possible hierarchical steps. 

Figure \ref{fig:chimdec2} depicts a similar stepwise transition from nonlocal to hierarchical coupling for the symmetric base pattern $(110011)$ with $n=4$, which has the same fractal dimension and link density but larger clustering coefficients. The total system size is $N=1297$ and the clustering coefficients are shown in Fig.~\ref{fig:cc1} (orange squares). For $m=1$ we again observe a $2-$chimera, like in Fig.~\ref{fig:chimdec1}(a). At $m=2$, the multiplicity of the chimera increases to a $7-$chimera which remains stable for further hierarchical steps, and is characterized by a more pronounced $\omega_k$ profile at the fully hierarchical level $m=n=4$.\\ 
The importance of the clustering coefficient as a measure for hierarchical systems with respect to chimera states is further highlighted by considering several permutations of this base pattern of length $b=6$, and $n=4$. We have performed a multitude of scans over wide ranges in ($\varepsilon,\sigma$)-space, for various less symmetric base patterns and varying initial conditions (specially prepared chimera-like as well as random), but none of them resulted in a chimera state. Among them are topologies with the same $c_1=4$ but different link distributions and lower clustering coefficients, such as $(100111)^4$, $(011101)^4$, $(110101)^4$, $(011110)^4$ and $(110110)^4$, but also systems with larger link densities and clustering coefficients $(011111)^4$, $(110111)^4$. Each base pattern at fixed hierarchical step has a fixed relative coupling radius, i.e., ratio between number of links for each element and size $N$ of the system. Usually, intermediate values of the relative coupling radius promote chimera states, whereas at the same time for small and large relative coupling radius chimeras are rarely observable. The subtle interplay of symmetry and compactness of the network topology with the number of links crucially affects the network dynamics. Moreover, due to multistability, the choice of initial conditions usually plays an important role. While it is impossible to draw a definite conclusion from purely empirical studies, this strongly suggests that generally the combination of a symmetric base pattern with large clustering coefficient promotes the existence of chimera states in hierarchical topologies. 
\begin{figure*}[t]%
\includegraphics[width=0.8\textwidth]{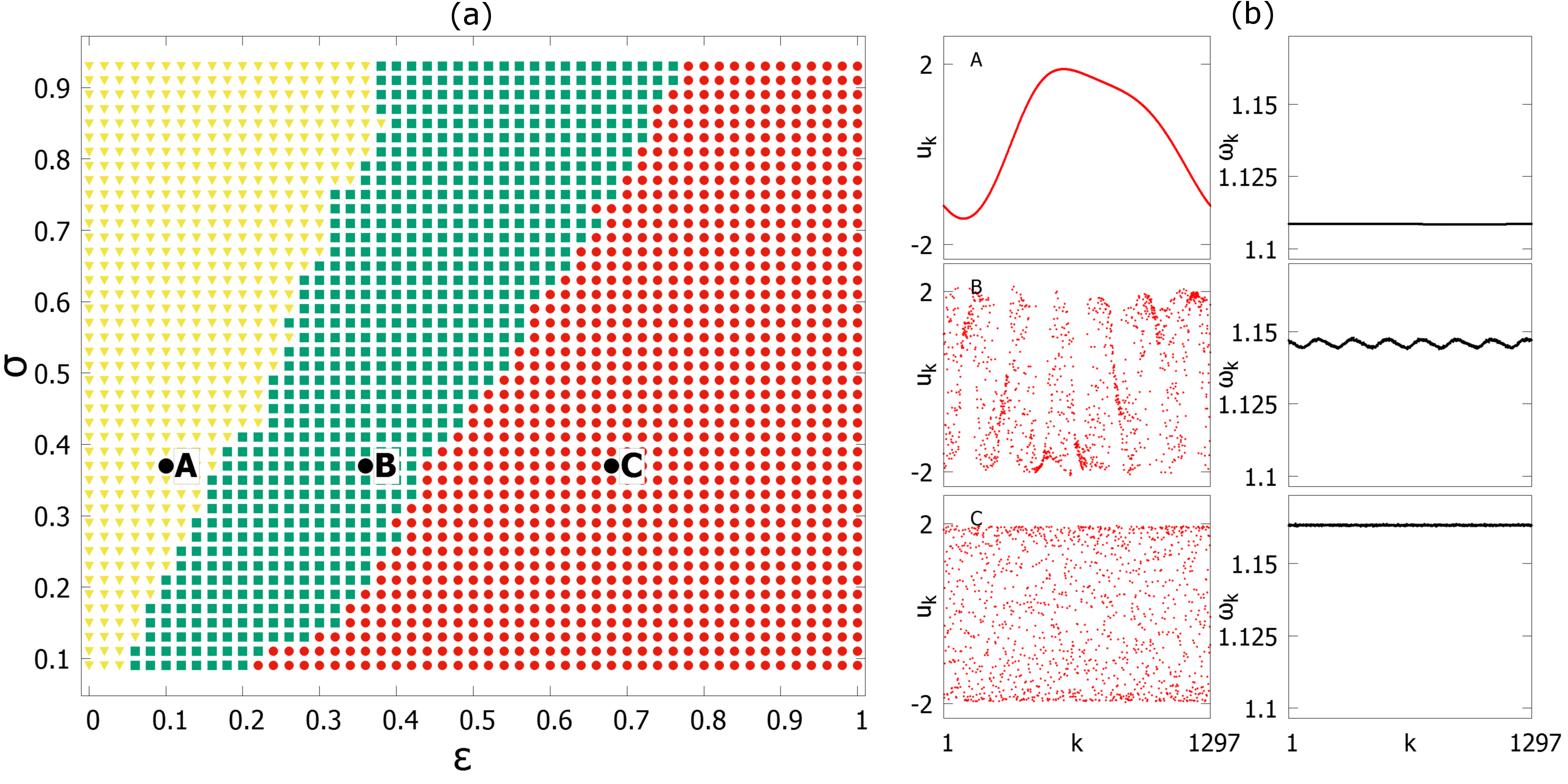}%
\caption{Stability regimes for the system $b_{init}=(110011)$ in an $m=n=4$ fully hierarchical topology, $N=1297$, $C(110011,4,4)=0.297791$, $\rho=0.1974$. (a) Diagram in the parameter space $(\varepsilon,\sigma)$: coherent state (yellow triangles),  $7-$chimera  state (green squares), incoherent state (red circles); (b) exemplary snapshots of variables~$u_k$ (left column) and mean phase velocities~$\omega_k$ (right column) for $\sigma=0.38$ and $\varepsilon=0.1$~(A), $\varepsilon=0.36$~(B), $\varepsilon=0.68$~(C).}
\label{fig:dynm4}%
\end{figure*}

\section{Stability regimes of chimeras in hierarchical networks}
\label{sec:regime}
To elaborate the role of the system parameters we construct the maps of stability regimes in the plane of the nonlinearity paremeter~$\varepsilon$ of the individial oscillators, and the coupling strength~$\sigma$. For this purpose we choose a system with base pattern $(110011)$ and $n=4$, and consider its $3-$step and $4-$step hierarchical topology.
We start from specially prepared chimera-like initial conditions for a fixed parameter set, and then use the obtained final state as an initial condition for the subsequent set of parameters, and so forth with a step size $\Delta \varepsilon = 0.02$ and $\Delta \sigma = 0.02$.

Fig.~\ref{fig:dynm3}(a) depicts the stability regimes for the base pattern $b=(110011)$ with $n=4$ in a $3$-step hierarchical topology ($N=1297$).
For small values of the nonlinearity parameter~$\varepsilon$ a completely coherent state with wave-like profile is observed (yellow triangles, and snapshot A in Fig.~\ref{fig:dynm3}(b)). Conversely, for larger $\varepsilon$, when the limit cycle of each individual oscillator starts to transform from sinusoidal to relaxation oscillations, we observe completely incoherent states (red circles, and snapshot C in Fig.~\ref{fig:dynm3}(b)). Between these two regimes, there is a region where chimera states emerge (green squares, and snapshot B in Fig.~\ref{fig:dynm3}(b)). Thus the transition from coherence to incoherence occurs through chimera states. 
Furthermore, the boundaries between the cohererent regime and the chimera state as well as between the chimera state and the completely incoherent regime shift to larger $\varepsilon$ with increasing $\sigma$. Eq.~(\ref{eqn:gen1}) shows that an upscaling of the local parameter $\varepsilon$ has to be counterbalanced by the coupling strength term which is controlled by $\sigma$ in order to lead to a similar force for each oscillator. Notably, the boundaries for chimera emergence in the ($\varepsilon,\sigma$) plane are approximately given by straight lines. In the right panel of Figure \ref{fig:dynm3}(b) the mean phase velocity profiles are shown for the three selected points A, B, C in the ($\varepsilon,\sigma$) plane. The chimera state in B shows chimera dynamics where the minima of the profile correspond to the coherent regions.

In contrast, Fig.~\ref{fig:dynm4}(a) shows the same dynamic regimes in ($\varepsilon,\sigma$) space for the fully hierarchical system 
($110011$) with $n=4$. The same qualitative shift and increase of the stable $\varepsilon$ range with increasing $\sigma$ is observed. However, the overall area of stability significantly decreases for the fully hierarchical system, while the character of the chimera state (multiplicity and coherent regions) remains unchanged. The system undergoes the same qualitative change of its dynamic behavior, a transition from a completely coherent state (yellow triangles) to a completely desynchronized state (red circles) via a 7-chimera (green squares) with increasing $\varepsilon$. The only difference between both systems is the decrease in the clustering coefficient $C$ and the link density $\rho$ with increasing hierarchical step $m$. This indicates that a high clustering coefficients and an increased number of links promote the existence as well as the stability of chimera states in hierarchical topologies. Stable chimera states are difficult to observe in hierarchical systems that have a very low clustering coefficient.

In this exemplary system, as discussed in the previous section, we have observed only chimera states with $7$ incoherent domains. For systems with nonlocal coupling topology we have shown recently that an appropriate choice of the coupling radius determines the multiplicity of the incoherent regions in chimera states, but there may be multistability between different multi-chimera states~\cite{OME15a}. Therefore, it is possible that with other initial conditions one can also find other multi-chimera states.

It is remarkable that the regime of chimera states in hier\-archical networks (Figs.~\ref{fig:dynm3}, \ref{fig:dynm4}) extends to much larger values of the coupling strength $\sigma$ than has been found for nonlocal coupling \cite{OME15a}. This indicates that quasi-fractal connectivities promote chimeras at large coupling strength where in more compact topologies completely coherent states prevail.

\section{Larger base pattern}
\label{sec:larger}
\begin{figure*}[t]%
\includegraphics[width=0.9\textwidth]{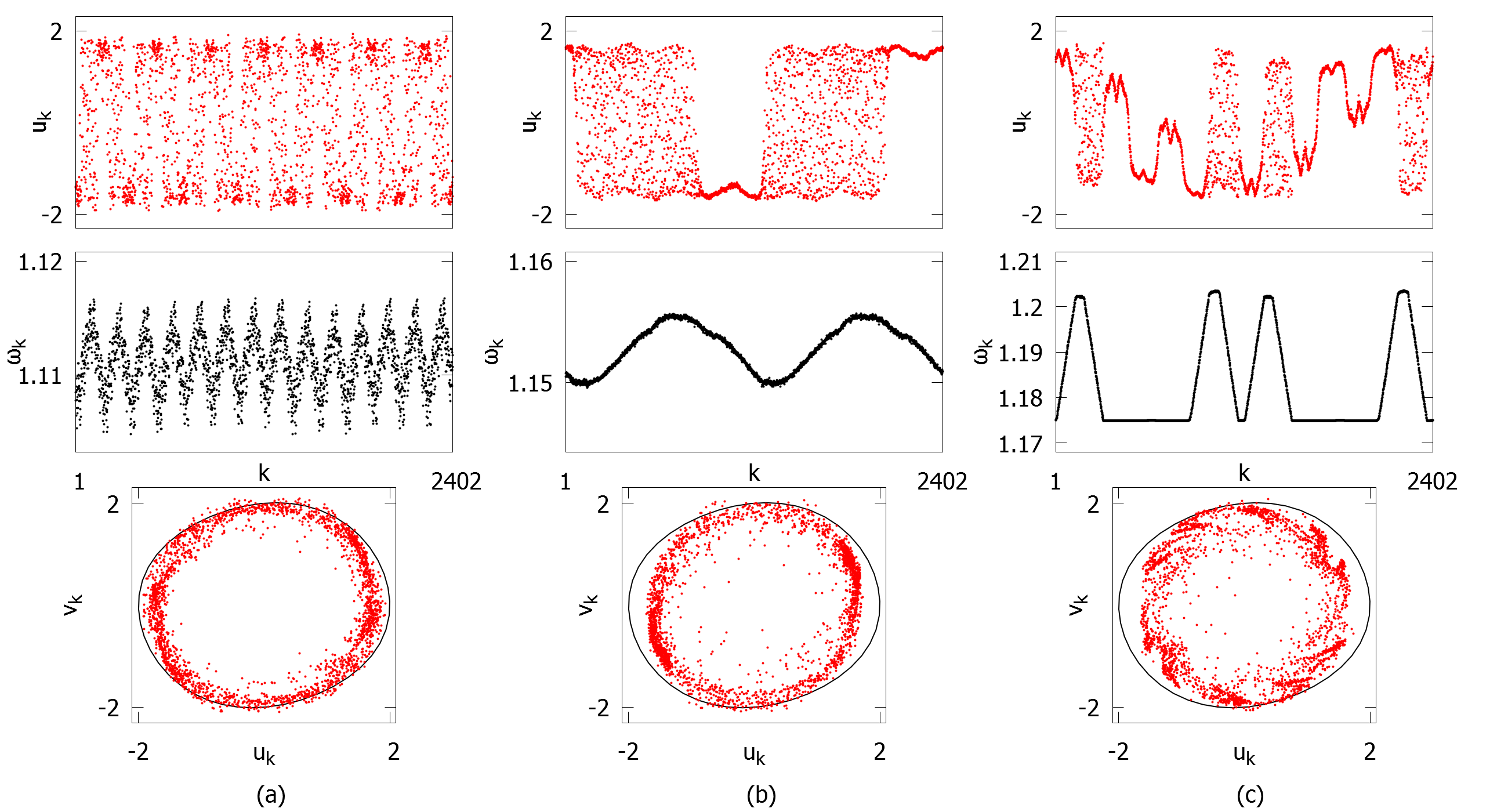}%
\caption{Snapshots of variables $u_k$ (upper panels), mean phase velocities $\omega_k$ (middle panels), and snapshots in the phase space $(u_k,v_k)$ (bottom panels, limit cycle of the uncoupled unit shown in black) for fully hierarchical system $b_{init}=(1110111)$, $n=4$ with $N=2402$, $C(1110111,4,4)=0.548829$, $\rho=0.54$, and $\varepsilon=0.1$, random initial conditions; (a)~$\sigma=0.25$, $14-$chimera; (b)~$\sigma=0.35$, $2-$chimera; (c)~$\sigma=0.45$, irregular $4-$chimera with nested regions of coherence. Note strong amplitude dynamics in all bottom panels.}%
\label{fig:1110111res}%
\end{figure*}
Changing the size, symmetry and number of links in the base pattern b$_{init}$ can lead to various completely different hierarchical topologies. This strongly influences the resulting clustering coefficient and the link density of the final system. As previous studies of chimera bifurcation scenarios\cite{OME15a} have shown, a change in the effective coupling radius of a nonlocal topology leads to different regimes of chimera states with various multiplicities. The same is true when the link density is increased. Since for the systems with $b=6$, $n=4$, $c_1=4$ we have only found chimera states in some symmetric configurations, we will now consider larger base patterns. This, consequently, allows for more links in the base pattern, and generates networks with higher clustering coefficients in the fully hierarchical system $m=n$. However, this dramatically increases the network size, making numerical simulations more expensive.

As an example we use the base pattern b$_{init}$=(1110111) with n=4 to generate a hierarchical topology of size N=2402 after full Cantor iteration. The link density is $\rho=\frac{g}{N}=0.54$ which is considerably larger than for the previously considered examples with base pattern of length $b=6$. Furthermore, the clustering coefficient of the fully hierarchical network $C(1110111,4,4)=0.548829$ is larger than in all previous examples.

We fix the nonlinearity parameter $\varepsilon=0.1$, and observe the system dynamics for three increasing values of coupling strength $\sigma=0.25, 0.35, 0.45$, starting from random initial conditions.  Fig.~\ref{fig:1110111res} depicts the corresponding snapshots, mean phase velocity profiles, and phase portraits. For small coupling strength (Fig.~\ref{fig:1110111res}(a)), a chimera state with high multiplicity ($14-$chimera) is obtained. With increasing coupling strength we move towards chimera states with two incoherent domains, but this chimera performs much stronger amplitude dynamics (bottom panel in Fig.~\ref{fig:1110111res}(b)). Such two different types of chimera states with strong and weak amplitude dynamics, respectively, were recently observed also in nonlocally coupled networks of Van der Pol oscillators~\cite{OME15a}.
Further increase of the coupling strength results in a complex 4-chimera state (Fig.~\ref{fig:1110111res}(c)). Here, the amplitude dynamics is even stronger and the coherent regions exhibit an additional substructure. In hierarchical systems of FitzHugh-Nagumo oscillators\cite{OME15}, nesting effects that appear somewhat similar to the observed structure in the amplitude have been observed in the mean phase velocity $\omega_k$. In Fig.~\ref{fig:1110111res}(c) however, the substructure is observed in the amplitude (top panel) and not in the mean phase velocity profile (middle panel). A closer look at the phase portrait (bottom panel) shows strong amplitude dynamics of a peculiar vortex-like structure, caused by the clustering of oscillators in the different coherent regions. Here, clusters of coherent nodes oscillate on smaller cycles in phase space while incoherent nodes roughly follow the limit cycle of the uncoupled system.

Thus, large hierarchical networks allow us to observe a variety of chimera states with either weak or strong amplitude dynamics.

\section{Conclusion}
\label{sec:outlook}

In the current study, we have analyzed chimera states in networks with hierarchical topologies. Using a modified iterative Cantor construction algorithm, the network topology is tuned stepwise between nonlocal and hierarchical. We have identified the clustering coefficient and symmetry properties of the base pattern as crucial factors in classifying different
topologies with respect to the occurrence of chimera states. We show that symmetric topologies with large clustering coefficients promote the emergence of chimera states, while they are suppressed by slight topological asymmetries or small clustering coefficients.
We have determined stability regimes in the plane of coupling strength and nonlinearity parameter of the individual
oscillator, which show that chimera states indeed appear on the transition scenario between complete coherence and incoherence. The analysis of an exemplary network with larger base pattern, resulting in larger clustering coefficient and more complex network structure, has revealed two different types of chimera states highlighting the increasing role of amplitude dynamics. 

\begin{acknowledgments}
We thank Astero Provata for useful discussions about hierarchical topologies. 
This work was supported by Deutsche Forschungsgemeinschaft in the framework of Collaborative Research Center SFB~910.
\end{acknowledgments}

%

\end{document}